\documentclass[a4paper]{jpconf}
\usepackage{orcidlink}
\usepackage{graphicx}
\usepackage{upgreek}
\usepackage{tikz}
\usepackage{subfigure}
\usepackage{xcolor}
\usepackage{bm}
\usepackage{amssymb}
\usepackage[numbers,square,sort&compress]{natbib}
\usetikzlibrary{external}

\begin{document}
\title{Influence of the Reynolds number from $\bm{Re_\tau= 150}$ to $\bm{210}$ on size-dependent bipolar charging}

\author{S Jantač$^1$, H Grosshans$^1$ $^2$}
\address{$^1$ Physikalisch-Technische Bundesanstalt (PTB), Braunschweig, Germany}
\address{$^2$ Otto von Guericke University of Magdeburg, Institute of Apparatus- and Environmental Technology, Magdeburg, Germany}
\ead{simon.jantac@ptb.de}



\begin{abstract}
We recently found wall-bounded turbulence to suppress and control bipolar triboelectric charging of particles of identical material.
This control is due to fluid modifying the motion of light particles.
Thus, the particles' charge distribution depends on their Stokes number distribution.
More specifically, fluid forces narrow the bandwidth of the charge distribution, and bipolar charging reduces dramatically. 
Consequently, not the smallest but mid-sized particles collect the most negative charge. 
However, the influence of the Reynolds number or particle concentration on bipolar charging of polydisperse particles is unknown.
This paper presents the charging simulations of same-material particles the in different wall-bounded flows.
In a comprehensive study, we vary the Reynolds number from $Re_\tau=$ $150$ to $210$ and the particle number density from $4 \times 10^9 \ \mathrm{m}^{-3}$ to $1 \times 10^{10} \ \mathrm{m}^{-3}$ to further explore the influence of the carrier flow on bipolar charging.
We model charge transfer based on the balance of transferable charge species.
Such species can represent adsorbed ions transferred during collisions or free electrons captured into a lower energy state on the other surface.
The turbulent flow is modeled via Direct Numerical Simulations (DNS) and is coupled to the particulate phase modeled via the Discrete Element Method (DEM).
Overall, our multiphysics approach couples the fluid dynamics, electric field, triboelectric charging, and particle momentum into one complex simulation.
\end{abstract}

\section{Introduction}
Triboelectric charging causes a build-up of charge in particulate systems.
In industrial environments, the excessive build-up causes unwanted agglomeration and wall-sheeting in the fluidized beds \cite{FOTOVAT2017303,Yoshimatsu20166261,SALAMA201321,Manafi2021109,Song201714716}, pneumatic conveyors \cite{GROSSHANS2023140918,Grosshans2022,Gro20d,XU2023104970,GROSSHANS2022117623}, or during powder handling in general \cite{ZHOU2021772}.
This is an especially severe problem in the pharmaceutical industry \cite{ZAFAR2018151,Alfano2023,ALFANO2021491,SUPUK2012427} because the formulation of powders is strictly given and often cannot be changed to reduce the charging.
Triboelectric charging also plays a vital role in many natural phenomena.
Like volcanic eruptions, where the charging is one of the contributors to the formation of lighting, and the charge build-up hinders the correct prediction of tephra behavior \cite{Cimarelli2022,James2008}.
It was shown that during planetary formation \cite{Kai2022}, the electrostatic forces play an essential role, and thus accurate estimation of the charge build-up is vital.

In all the above-mentioned processes, the charge is transported due to a combination of inter-particle and particle-wall collisions.
The particle-wall collisions typically occur between objects of different materials, a situation that can be described by models that use a difference in the material property as a driving force for the charge transfer \cite{Chowdhury2021104}.
However, inter-particle collisions often occur between particles of identical material; thus, those models cannot describe experimentally observed bipolar charging in polydispersed systems of identical material \cite{Forward2009phys,KONOPKA2020713}.
New mosaic-based models were proposed that can explain those experiments based on the statistical variation of properties on otherwise identical materials \cite{Lacks2008,Baytekin2011}.
However, they describe the charging only at the nano/mesoscale.

Recently, we expanded our research to encompass turbulent flows with the bipolar charging of identical materials and unveiled how turbulence fundamentally alters the charge distribution  \cite{jantač2023suppression}. Our investigation revealed a major shift in highly polydisperse systems, specifically when the smallest particles possess a Stokes number of $ \approx 1.5$ or less. Surprisingly, in such cases, the mid-sized particles carry the most negative charge, while the smallest particles exhibit a negligible charge, a departure from the expected behavior. Consequently, this significantly narrows down the charge distribution in the polydisperse systems. Our analysis indicates that turbophoresis is responsible for this reduction, arising from the decreased local polydispersity within the system. Although our initial study was limited to specific flow conditions, we have now demonstrated that these earlier findings extend to other flow scenarios as well.

\section{Simulation setup}
Our simulation couples the DNS with the DEM approach to accurately simulate the particle trajectories. The fluid phase is fully resolved DNS simulation which is coupled with the particle phase \cite{OZLER2023103951,Grosshans2017465} . The particle phase is simulated via DEM, and we consider the following forces: The electrostatic force between all particles and grounded walls of the channel, the drag force, and the lift force in the turbulent channel.

The domain was set up as a channel flow, i.e., boundaries of the system in the $y$-axis were used as walls (zero slip conditions), the boundaries in the $z$-axis were set to be periodic, and boundaries in the $x$-axis were set as periodic.
The dimension of the domain was $6H \times H \times 2H$ where $H=4$ cm [see Fig. \ref{fig:DNS}]. The corresponding grid resolution was $256 \times 144 \times 144$, and the size of the cells in the $x$ and $z$ axis is uniform.
This domain setup mimics a section of the pneumatic conveyors.

\begin{figure}
    \centering
    \includegraphics{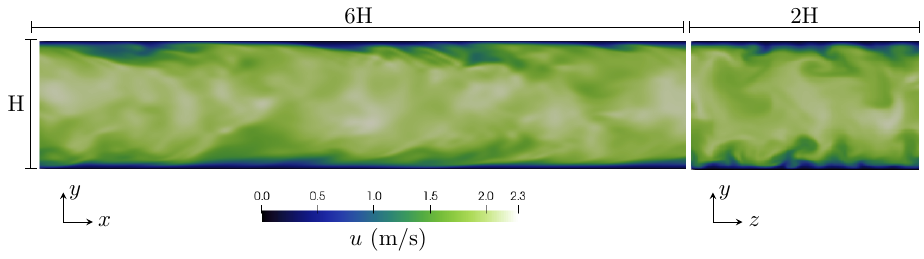}
    


    
    \label{fig:DNS}
    \caption{DNS simulation of a channel flow. On the left is the side view where the airflow is traveling from left to right. On the right is the front view of the simulation domain.}
    \label{fig:DNS}
\end{figure}

Our charging model is based on the balance of charge carriers initially homogeneously distributed on the surface.
Lacks formulated this model type \cite{Lacks2008}, and Grosjean formulated a newer discretized version \cite{Grosjean2020,Grosjean23b}.
However, we will use a generalized version of those models formulated by Konopka and Kosek \cite{KONOPKA2017150}.

\begin{figure}
    \centering
    \subfigure[]{
    \includegraphics[width=0.48\textwidth]{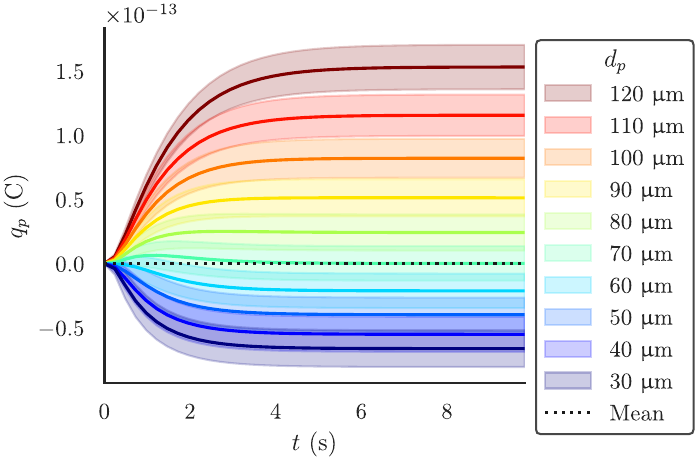}
    \label{fig:Averages_charge}
    }
    \subfigure[]{
    \includegraphics[width=0.48\textwidth]{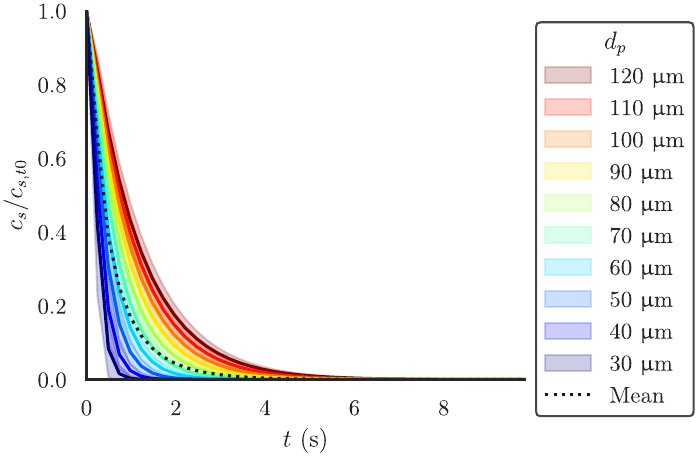}
    \label{fig:Averages_spec}
    }
    \caption{Typical bipolar charging of the polydisperse system in the absence of ambient flow. The color represents that particle size. a) The largest particles are most positively charged, and the smallest are most negatively charged. The net charge in the simulation is conserved and thus is zero through the simulation (see the dashed line). b) The charging rate depends on the charge carriers' concentration; the charging ceases when the charge carriers are depleted}.
    \label{fig:DEM}
\end{figure}

We assume that the surface consists of two types, one is the charge carrier, and the second one is the free spot where the charge carrier can be spontaneously transferred (relaxed).
The surface resistivity prevents any movement of the charge carrier.
Thus, the charge carrier can be transported and subsequently relaxed only when it collides with the free spot on the other surface.
Once the charge carrier is transported, we assume it is stable and does not participate in subsequent collisions.
Additionally, we assume that the concentration of transferable charge carriers is much lower than that of free spots.
This means that we can formulate a balance equation of only charge carriers to describe the net charge transfer during collision:
\begin{equation}
\Delta q_i = z e (c_{\mathrm{s},j}-c_{\mathrm{s},i})A_\mathrm{c,max}^{i,j}, 
\label{eq:balance}
\end{equation}
where $z$ is the charge number of the charge species, $e$ is the elementary charge, $c_{\mathrm{s},j}$ is the concentration of charge species on $j$-th particle, and finally $A_\mathrm{c,max}^{i,j}$ is the maximal collision contact area of $i$-th and $j$-th particle, which can be calculated as follows \cite{KOREVAAR2014144}: 
\begin{equation}
    A_\mathrm{c,max}^{i,j}=\pi \left(\frac{5 m_{i,j} r_{p,i,j}^2}{4 E_{i,j}}\right)^{2/5} v_{n}^{4/5},
\end{equation}
where $m_{i,j}$ is the effective mass, $r_{p,i,j}$ is the effective radius, and $E_{i,j}$ is the effective elastic modulus of colliding particles, which are calculated as the reciprocal value of the harmonic sum of $i$-th and $j$-th particles properties.
By balancing the charge carriers on the particle level, we additionally assume that enough time has passed for the particle to be randomly reoriented for the subsequent collision; this allows us to describe the charge transfer only as a function of charge carrier concentration.

Summing the charge transfer over $n$ collisions yields a net charge on each particle: 
\begin{equation}
    q_i=q_{i,t_0}+\sum_{i=1}^{n} \Delta q_i(n),
    \label{eq:netCharge}
\end{equation}
where $q_{i,t_0}$ is the initial charge of particle. 
To illustrate the charging of the polydisperse system with the above-mentioned model, we show the result of simulations in the absence of flow (Figure \ref{fig:DEM}).
The absence of flow simplifies the description and allows validation of our simulations with previously published works \cite{KONOPKA2017150,GE2023118180,YU2017113}.
We assume that particles are initially uncharged and have uniform concentrations of charge carriers.
This causes the initial few collisions of particles to transfer little charge because an equal amount of charge carriers are transported between colliding particles.
However, as charge carriers start to deplete, the rate of charging increases, and since small particles deplete the charge carriers faster than large ones [see Figure \ref{fig:Averages_spec}], they start to charge negatively.
In contrast, large particles charge positively because they are the donors of the negatively charged charge carriers. Electrons or adsorbed ions are examples of negatively charged carriers.

\section{Results}

Before we enabled the charging of the particles, we let both the fluid field and particle concentration fully converge.
After convergence, we initialized the triboelectric model.
The initial values of particles were set as follows: zero initial charge ($q_{i,t_0}=0$ C ), uniform charge carrier density ($c_{\mathrm{s},t_0}=10 \ \mathrm{\upmu m}^{-2}$), the charge number of the carrier was assumed to be equivalent to electron ($z=-1$), the elastic modulus of particles was set to $E=1 \times 10^8 \ \mathrm{Pa}$.

The charge build-up is shown in Fig. \ref{fig:Averages_charge}. The largest particles tend to charge positively.
And small particles tend to charge negatively. The decay of charge carriers has an exponential trend, as shown in Fig. \ref{fig:Averages_spec}.
However, as shown in our recent paper \cite{jantač2023suppression}, the charge distribution is qualitatively changed when the coupling between the fluid and particles is considered.
Here we study the effect of friction Reynolds number and particle number density.
In the flowing studies, we evaluate the charge distribution when the average concentration of charge species decreases by a characteristic value of $1/\mathrm{e}$ of initial concentration (here, the $\mathrm{e}$ is the Euler number).

\subsection{Friction Reynolds number}

\begin{figure}
    \centering
    \subfigure[]{
    \includegraphics[width=0.48\textwidth]{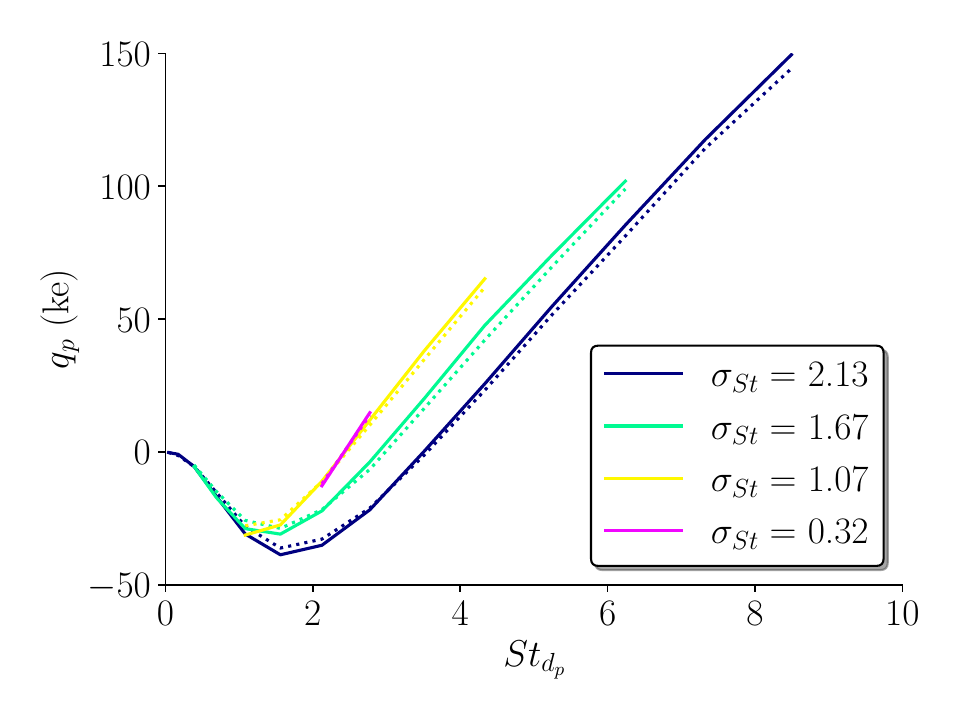}
    \label{fig:Chargedist}
    }
    \subfigure[]{
    \includegraphics[width=0.48\textwidth]{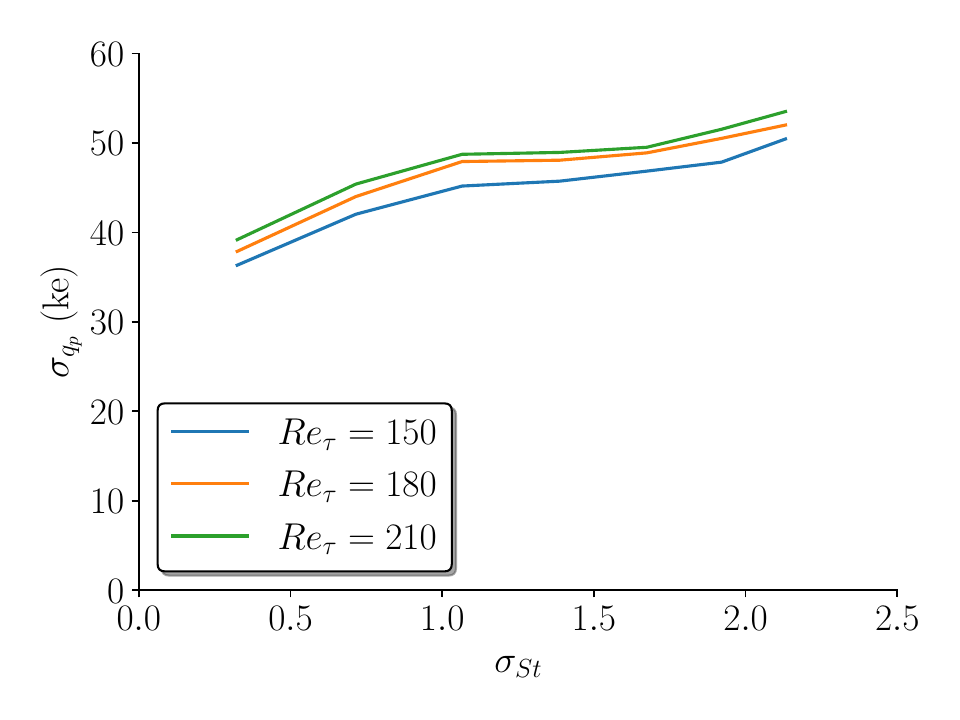}
    \label{fig:Re_tau}
    }
    \caption{a) The bipolar charging is significantly suppressed when $Re_\tau$ is varied from 150 to 210. The full lines denote the results when $Re_\tau=210$, and the dashed lines when $Re_\tau=150$. For clarity, we show every second case of $\sigma_{St}$, also note that the y-axis is scaled in the units of the elementary charge. b) the standard deviation of charge as a function of particle size distribution measured as a $\sigma_{St}$ shows an increase in the width of the charge distribution with an increase in $Re_\tau$. Values of $Re_\tau=180$ are taken from our recent paper \cite{jantač2023suppression}.}
    \label{fig:Re}
\end{figure}

In this section, we studied the effect of the friction Reynolds number on bipolar charging. We set the particle number density to be $4 \times 10^9 \ \mathrm{m}^{-3}$, and varied $Re_\tau$ from 150 to 210.
Additionally, we varied the particle size distribution from monodisperse to polydisperse, where the smallest particles were 10~\textmu m in diameter and the largest 140~\textmu m in diameter.
The results of those simulations show that in the studied range of $Re_\tau$, the qualitative shape of the charge distribution remains the same [see, Fig. \ref{fig:Chargedist}]. However slight increase in the width of the charge distribution was observed.
To quantify the width of the charge distribution, we evaluated the standard deviation of the charge on particles $\sigma_{q_p}$ and a standard deviation of the Stokes number of particles in the system $\sigma_{St}$.
The relative difference between the standard deviation of particle charge in all particle size distributions was around 7 \% [see Fig. \ref{fig:Re_tau}].In fact, the values of $\sigma_{q_p}$ appear to be proportionally shifted with increasing $Re_\tau$.
Such a consistent shift indicates that the distribution of inter-particle collisions is very similar in both cases, only the frequency of collision increases.

We confirmed that the bipolar charging of identical materials is suppressed in the studied range of $Re_\tau$ by measuring the width of the charge distribution at the evaluation point $c_{\mathrm{s}}=c_{\mathrm{s},t_0}/\mathrm{e}$, but the time required to reach this point was reduced from $\approx 3.6$ s when  $Re_\tau = 150$ to $\approx 3.0$ s when  $Re_\tau = 210$.

\subsection{Particle number density}
\begin{figure}
    \centering
    \subfigure[]{
    \includegraphics[width=0.48\textwidth]{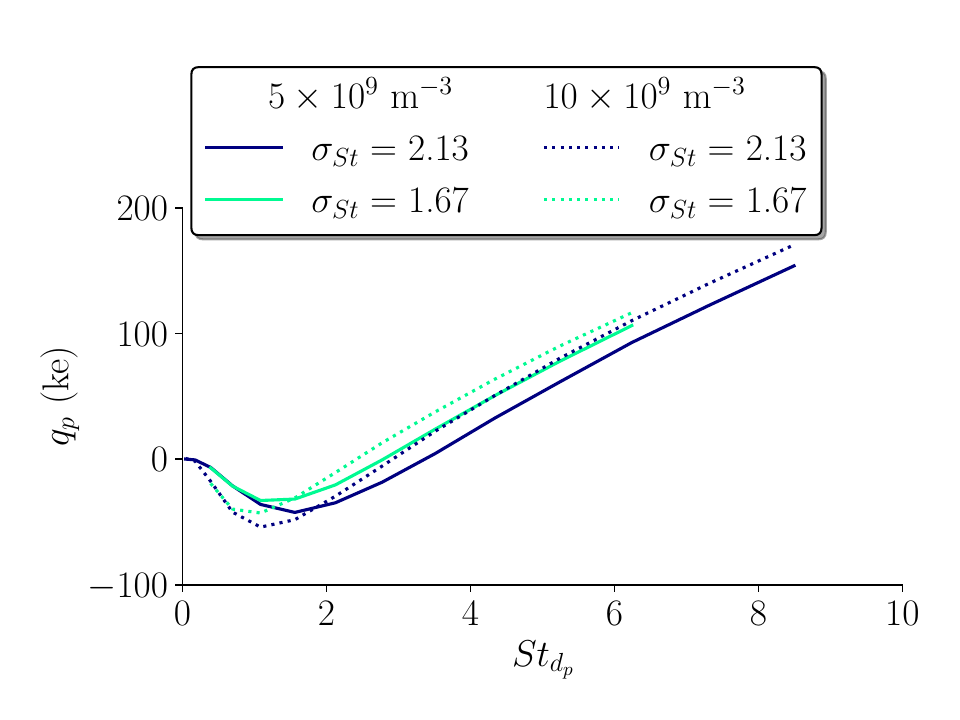}
    \label{fig:Chargedist_vol}
    }
    \subfigure[]{
    \includegraphics[width=0.48\textwidth]{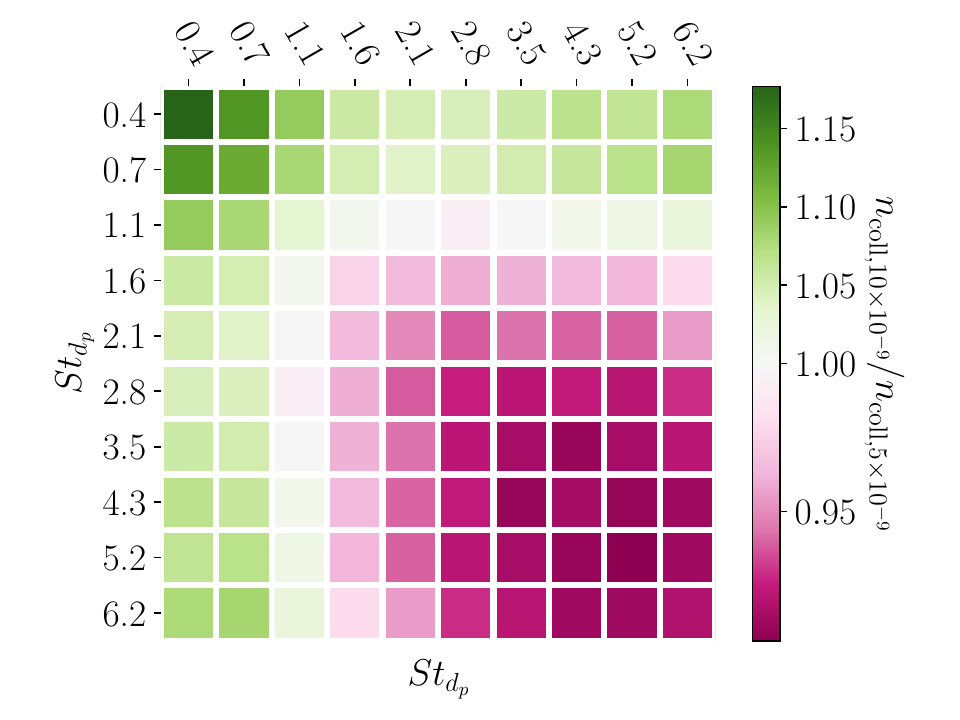}
    \label{fig:Collisions}
    }
    \caption{a) The charge distribution for two most polydisperse cases, when $Re_\tau=210$. The increase in volumetric density causes increased charge accumulation. b) The distribution of inter-particle collisions for $\sigma_{St}=1.67$ shows that in denser cases, the smaller particles collide more frequently with particles of all sizes, and large particles collide less frequently with particles of similar size.}
    \label{fig:Re_vol}
\end{figure}

The particle number density plays a stronger role in bipolar charging. Studying two cases of the particle number density of $5 \times 10^{-9}\ \mathrm{m^{-3}}$, and $10 \times 10^{-9}\ \mathrm{m^{-3}}$, showed that the charge distribution is significantly changed [see Fig. \ref{fig:Chargedist_vol}]. Neutral particles are smaller when the number density is larger. The charge of most negative particles is higher for higher number density. This indicates that inter-particle collisions are significantly changed [see Fig. \ref{fig:Collisions}]. In the denser case, the particles of $St_{d_p}=0.4$ and $St_{d_p}=0.7$ collided more frequently with all particles; this causes them to lose charge carriers faster than other particles, making them negatively charged. The average collision ratio is similar for both cases for particles of $St_{d_p}=1.6$ which makes them similarly charged [see the intersection of green lines at Fig. \ref{fig:Chargedist_vol}].
A particle with $St_{d_p} \ge 2.1$ collided more frequently with smaller particles and less frequently with larger particles. Since in later stages of charging, the smaller particles have a lower concentration of charge carriers than large particles [see Fig. \ref{fig:Averages_spec}], the large particles in denser cases charged more positively.
This indicates that bipolar charging will be more significant in systems with high particle number density.

\section{Conclusions}
Triboelectric charging causes severe problems in industry and frequently occurs in nature.
Current triboelectric models are based on the nano/mesoscale, which limits their prediction to, at best, two-particle charging and naturally cannot describe emergent properties. However, the above-mentioned phenomena are macroscopic and therefore include a vast number of particles whose collective charging behavior cannot be described simply by fundamental models. To improve our understanding of such a system, we simulate a large-scale powder system in a change flow, where particles of identical material properties charge bipolarly based on recent mosaic models. Our simulation couples DNS and DEM in a four-way coupling which includes the coupling of the gas and particle phase, the particle position is coupled with the electrostatic forces created by the accumulated charge, and lastly, the particle charge is dependent on the particle collisions and positions.

Our research has revealed that turbulence is crucial in suppressing bipolar charging across a broader spectrum of flow conditions and particle number densities. Specifically, the width of the charge distribution is influenced by the friction Reynolds number, but it doesn't significantly alter the previously reported qualitative change in the charge distribution as described in \cite{jantač2023suppression}. As the Reynolds number increases, collision frequency rises proportionally, leading to a shift in particle charge standard deviation. However, the particle number density has a more pronounced impact on bipolar charging. With higher particle number densities, the charge distribution undergoes more substantial changes, causing smaller particles to charge negatively and larger particles to charge positively due to differences in collision frequencies.
Overall, our findings demonstrate that in turbulent environments, bipolar charging experiences suppression at least within the range of $Re_\tau$ 150 to 210 and at particle number densities between $5 \times 10^{-9}\ \mathrm{m^{-3}}$ and $10 \times 10^{-9}\ \mathrm{m^{-3}}$.

The findings suggest that understanding and managing triboelectric charging are crucial for optimizing particulate systems in industrial processes. Proper consideration of charging effects can lead to better control and design of such systems. Our conclusions can be extended to polydisperse turbulent flows inside volcanic plumes or dust storms.

 \ack
 This project has received funding from the European Research Council~(ERC) under the European Union's Horizon 2020 research and innovation programme~(grant agreement No.~947606 PowFEct).

\bibliography{References}
\end{document}